\documentclass[a4paper]{amsart}

\usepackage{amsmath, amssymb, graphicx}  
\usepackage[utf8]{inputenc}

\newcommand{\lb}{\left(}
\newcommand{\lsb}{\left[}
\newcommand{\lcb}{\left\{}
\newcommand{\rb}{\right)}
\newcommand{\rsb}{\right]}
\newcommand{\rcb}{\right\}}
\newcommand{\lpm}{\begin{pmatrix}}
\newcommand{\rpm}{\end{pmatrix}}
\newcommand{\abs}[1]{\lvert #1 \rvert}
\DeclareMathOperator{\pder}{\partial}

\newcommand{\calF}{\mathcal{F}}
\newcommand{\calP}{\mathcal{P}}
\newcommand{\calV}{\mathcal{V}}
\newcommand{\calR}{\mathcal{R}}

\begin{document}

\title{Analysis of an SIR--model with global and local infections}
\author{Thomas Götz}
\email{goetz@uni-koblenz.de}
\address{Mathematical Institute, University Koblenz, D--56070 Koblenz, Germany}


\begin{abstract}
An epidemic model where disease transmission can occur either through global contacts or through local, nearest neighbor interactions is considered. The classical SIR--model describing the global interactions is extended by adding additional equations for the density of local pairs in different epidemic states. A locality parameter $p\in [0,1]$ characterizes the probability of global or local infections. The equilibria of the resulting model are analyzed in dependence of the locality parameter and the transmission rate of the pathogen. An explicit expression for the reproduction number in terms of the locality parameter and the disease parameters is obtained. Transient simulations confirm these findings. Neighboring pairs of one infected and one susceptible can be considered as active pairs, since local transmission of the disease can only occur in that situation. Our analysis shows, that the fraction of active pairs is minimal for intermediate values of the locality parameter. 
\end{abstract}

\maketitle

\section{Introduction}
Compartmental models to describe the dynamics of diseases have been analyzed since decades, starting with the works by Kermack--McKendrick~\cite{KerMcK27}. Classical SIR models assume a homogeneous mixing of susceptible, infected and recovered individuals, such that each infected can transmit the disease to any susceptible with the same rate. However network effects~\cite{Hesea22, Wijea21} or spatial inhomogeneity~\cite{Engea20, Vigea21} are known to have important effect on the transmission dynamics. For the current SARS--COV--2 pandemic, household effects have been identified as a prominent example of transmission networks~\cite{Madea20, Hesea22}. These network effects can be viewed as \emph{local} transmissions in contrast to the \emph{global} transmissions that are included in the classic SIR--model or its variants with refined compartmental structure. 

Spatial PDEs, network--based models or stochastic simulations are typical approaches to combine local and global transmission mechanisms. Cellular automata \cite{Matea92, DotFab15, Keeea16} can consider the global, long--range infection of a susceptible by any arbitrary infected as well as local, short--ranged infections due to neighboring infected individuals. In the sequel, we will consider the transmission rate as a mix of the global transmission $p\beta$ and the local transmission $(1-p)\beta$ between the four neighbors on a square lattice. The probability $p\in [0,1]$ can be viewed as a \emph{locality parameter} modeling the ratio of global transmissions compared to the overall transmissions. The case $p=1$, i.e.~only global transmissions and no local transmissions corresponds to the classical spatial homogeneous SIR--model. The other extremal case $p=0$ describes only local and no global transmissions where infections occur only between neighboring individuals. 

To mitigate the spread of epidemics, non--pharmaceutical interventions (NPI) are a frequently used measure~\cite{Fer20, Bocea20}. Contact restrictions and lock--downs as they are imposed during the SARS--COV--2 pandemic almost all over the world, can be viewed as a shift in the locality parameter $p$. The stricter the lock--down, the less likely global infections should be. However, the local infections occurring inside households and other closely related groups are still continuing and rather insensitive to government--imposed measures. Hence, analyzing the behavior of an extended SIR--model including both, global and local infections, can provide further insight into the effect of NPI's when altering the ratio between global and local transmissions.

Besides the locality parameter $p$, the transmission rate $\beta$ of the pathogen plays an important role. During the ongoing pandemic, several variants of the SARS--COV--2 virus appeared like the \emph{alpha}--, \emph{delta}-- and \emph{omicron}--subtype. Concerning their transmissibilty, \emph{omicron} seems to have a three--times higher transmission rate than the previously dominant \emph{delta}--strain~\cite{Itiea21}. On the other hand, the recovery rate seems to be comparable for both strains, see e.g.~\cite{Weeea22}. Therefore we will use the transmission rate $\beta$ as varying parameter in our analysis to illustrate the sensitivity of equilibria with respect to varying transmissibility of the pathogen.

The paper is organized as follows: In Section~\ref{S:Model} we recall a pair--approximation model~\cite{MalFab16, WreBes21, JooLeb04} which can be viewed as an extension of the classical SIR--model~\cite{KerMcK27, Mar15}. The \emph{density} of pairs $SS, SI, \dots$ of two susceptible, one susceptible and one infected individual, etc. are also taken into account. Depending on the locality parameter $p \in [0,1]$ our model switches between the SIR--model ($p=1$) and the pure pair approximation ($p=0$). To gain insight into the behavior of the mixed model $0<p<1$, we investigate its equilibria in Section~\ref{S:Equilibria}. We compare the known results for the SIR--model with findings for the local and mixed model. The next generation matrix approach allows us to compute the basic reproduction number for the mixed model in Section~\ref{S:R0}. Transient simulations of the mixed model are presented in Section~\ref{S:transient} to illustrate the convergence to the respective equilibria for various parameter regimes.

\section{SIR--Model with Global and Local Infections}
\label{S:Model}

We consider the global--local SIR model with pair approximation as presented by Maltz and Fabricius~\cite{MalFab16}. Introducing the normalized susceptible, infected and recovered compartments $S,I,R$, pairs of two individuals can be in one of the following six states: $SS$, $SI$, $SR$, $II$, $IR$ and $RR$. Let $Z_{ij}$ denote the normalized number of pairs in state $ij$. Assuming a constant normalized population $N=1$, it holds that $S+I+R=1$ as well as $Z_{SS}+Z_{SI}+Z_{SR}+Z_{II}+Z_{IR}+Z_{RR}=2$, since there are in total $2N$ pairs. Following the derivation presented in~\cite{MalFab16} we end up with an ODE--system taking into account the susceptible and infected compartments $S,I$ as well as the $SS$ and $SI$--pairs denoted by $X=Z_{SS}$ and $Y=Z_{SI}$.
\begin{subequations}
\label{E:Pairmodel}
\begin{align}
	S' &= \mu(1-S) - p \beta S I - \frac{q \beta}{4} Y \label{E:1}\\
	I' &= p \beta SI + \frac{q \beta}{4} Y - (\gamma+\mu) I \label{E:2}\\
	X' &= \mu (4S-2X) - 2 p\beta XI - \frac{3}{8} q \beta \frac{XY}{S}\label{E:3} \\
	Y' &= \mu (4I-Y) + p\beta (2X-Y) I + \frac{3}{8} q\beta \frac{XY}{S} 
		- \frac{q \beta}{4} Y\lb \frac{3}{4} \frac{Y}{S}+1\rb - (\gamma+\mu)Y  \label{E:4}
\end{align}
\end{subequations}
The remaining compartments $Z_{SR}, Z_{II}, Z_{IR}$ and $Z_{RR}$ are decoupled from this four--dimensional system. The transmission rate is denoted by $\beta$, the recovery rate equals $\gamma$ and $\mu$ denotes the birth and death rates. By $p$ we denote the probability for a global transmission and $q=1-p$ describes the probability of a local transmission. In case of $p=1$ and $q=0$, we obtain the standard Kermack--McKendrick SIR--model~\cite{KerMcK27} where only global infections occur. In the other extremal case $p=0$ and $q=1$ we obtain the pair approximation model presented by Joo and Lebowitz~\cite{JooLeb04}. 

For later reference we state the Jacobian of the system~\eqref{E:Pairmodel} depending on the locality parameter $p$
\begin{equation}
\label{E:J_p}
\begin{tiny}
	J_p = \lpm -\mu-p\beta I & -p\beta S & 0 & -q\beta/4 \\
		p\beta I & p\beta S-(\gamma+\mu) & 0 & q\beta/4 \\
		4\mu+\frac{3q\beta XY}{8S^2} & -2p\beta X & -2\mu-2p\beta I-\frac{3q\beta Y}{8S} & -\frac{3q\beta X}{8S} \\
		-\frac{3q\beta Y}{16S^2}(2X-Y) & 4\mu+p\beta (2X-Y) & 2p\beta I+\frac{3q\beta Y}{8S} &
		-(\gamma+2\mu+\tfrac{q\beta}{4}) -p\beta I +\frac{3q\beta (X-Y)}{8S} \rpm\;.
	\end{tiny}
\end{equation}

\section{Equilibria}
\label{S:Equilibria}
To determine equilibria of the above model~\eqref{E:Pairmodel}, we consider its stationary state, i.e.~$(S,I,X,Y)'=0$. Adding Eqns.~\eqref{E:1} and~\eqref{E:2} we can solve for $I$ and get
\begin{align}
	I &= I(S) = \frac{\mu}{\gamma+\mu}(1-S)\;.
\intertext{In case of $q\neq 0$, we use Eqn.~\eqref{E:2} to determine}
	Y &= Y(S) = \frac{4\mu}{q\beta} \lb 1- \frac{p\beta}{\gamma+\mu}S\rb\, \lb 1-S\rb\;.
\end{align}
In the local situation $p=0$ and $q=1$, the above equation simplifies to
\begin{equation*}
	Y_0 = Y_0(S) = \frac{4\mu}{\beta} (1-S)\;.
\end{equation*}
Next, we use~\eqref{E:3} to determine $X$ as a function of $S$
\begin{gather}
	X \cdot \lb 2\mu+2 p\beta I + \frac{3q\beta}{8}\frac{Y}{S}\rb = 4\mu S \notag
\intertext{and hence}
\label{E:XofS}
	X = X(S) =\frac{8 S^2}{S+3+\frac{p\beta}{\gamma+\mu} S(1-S)}\;.
\end{gather}
In the purely local case $p=0$, we get $X_0(S) = 8S^2/(S+3)$ which is completely independent of $\beta, \mu$ and $\gamma$. Finally Eqn.~\eqref{E:4} yields an equation for S
\begin{equation}
\label{E:Fp}
	0 \stackrel{!}{=} F_p(S) := \mu(4I-2Y) + p\beta (2X-Y) I + \frac{3q\beta}{16} \frac{2X-Y}{S}Y 
		- \lb \frac{q\beta}{4}+\gamma \rb Y \;.
\end{equation}
Since both $I$ and $Y$ contain a factor $\mu(1-S)$, we may write 
\begin{equation}
\label{E:Gp}
	F_p(S) := \mu(1-S)\cdot G_p(S)\;,
\end{equation}
this shows, that $S=1$ is the trivial, disease--free equilibrium $(S^0,I^0,X^0,Y^0)=(1,0,2,0)$. Other equilibria must be zeros of the function $G_p$. Before analyzing the function $G_p$ in order to determine the equilibria for the general model, we consider first the two special cases $p=1$ (classical SIR--model) and $p=0$ (only local infections).

\subsection{Global Model $p=1$}
The case $p=1$ and $q=0$ corresponds to the classical Kermack--McKendrick equations
\begin{align*}
	S' &= \mu(1-S) -\beta S I \\
	I' &= \beta S I - (\gamma+\mu) I
\end{align*}
In this case, the equation $F_1(S) = \mu(1-S)\lsb \frac{\beta S}{\gamma+\mu} - 1\rsb = 0$ determines the equilibria. Following the above notation in Eqn.~\eqref{E:Gp}, the function $G_1$ is given by
\begin{equation*}
	G_1(S) = \frac{\beta S}{\gamma+\mu} - 1\;.
\end{equation*}
The endemic equilibrium $S_1^\ast=\frac{\gamma+\mu}{\beta}$ corresponds to $I_1^\ast = \frac{\mu}{\gamma+\mu}-\frac{\mu}{\beta}$. For $\beta>\tilde{\beta}_1=\gamma+\mu$ the endemic equilibrium exists and can be shown to be asymptotically stable. Two other compartments $X$ and $Y$ read in the endemic equilibrium as 
\begin{equation*}
	X_1^\ast = 2{S_1^\ast}^2 = \frac{2(\gamma+\mu)^2}{\beta^2} \quad \text{and}\quad
	Y_1^\ast =  \frac{4\mu+2\beta X_1^\ast}{\gamma+2\mu+\beta I_1^\ast} I_1^\ast
		= 4 S_1^\ast I_1^\ast\;.
\end{equation*}

\subsection{Local Model $p=0$}
In the local model $p=0$ and $q=1$, we summarize
\begin{equation*}
	I_0 = \frac{\mu}{\gamma+\mu}(1-S)\;, \quad
	Y_0 = \frac{4\mu}{\beta} (1-S)\;, \quad
	X_0 = \frac{8S^2}{S+3}
\end{equation*}
and
\begin{align*}
	F_0(S) &= \mu(4I-2Y) 
	+ \lsb \frac{3\beta}{16}\frac{2X-Y}{S} -\frac{\beta}{4}-\gamma \rsb Y\;.
\intertext{Factoring out $\mu(1-S)$ yields}
	G_0(S) &= \frac{12S}{S+3} - \frac{3\mu}{\beta S} + C_0\;,
\end{align*}
where $C_0 = \frac{4\mu}{\gamma+\mu}-\frac{4\gamma+5\mu}{\beta}-1$.

\begin{figure}[htb]
\centerline{\includegraphics[width=.67\textwidth]{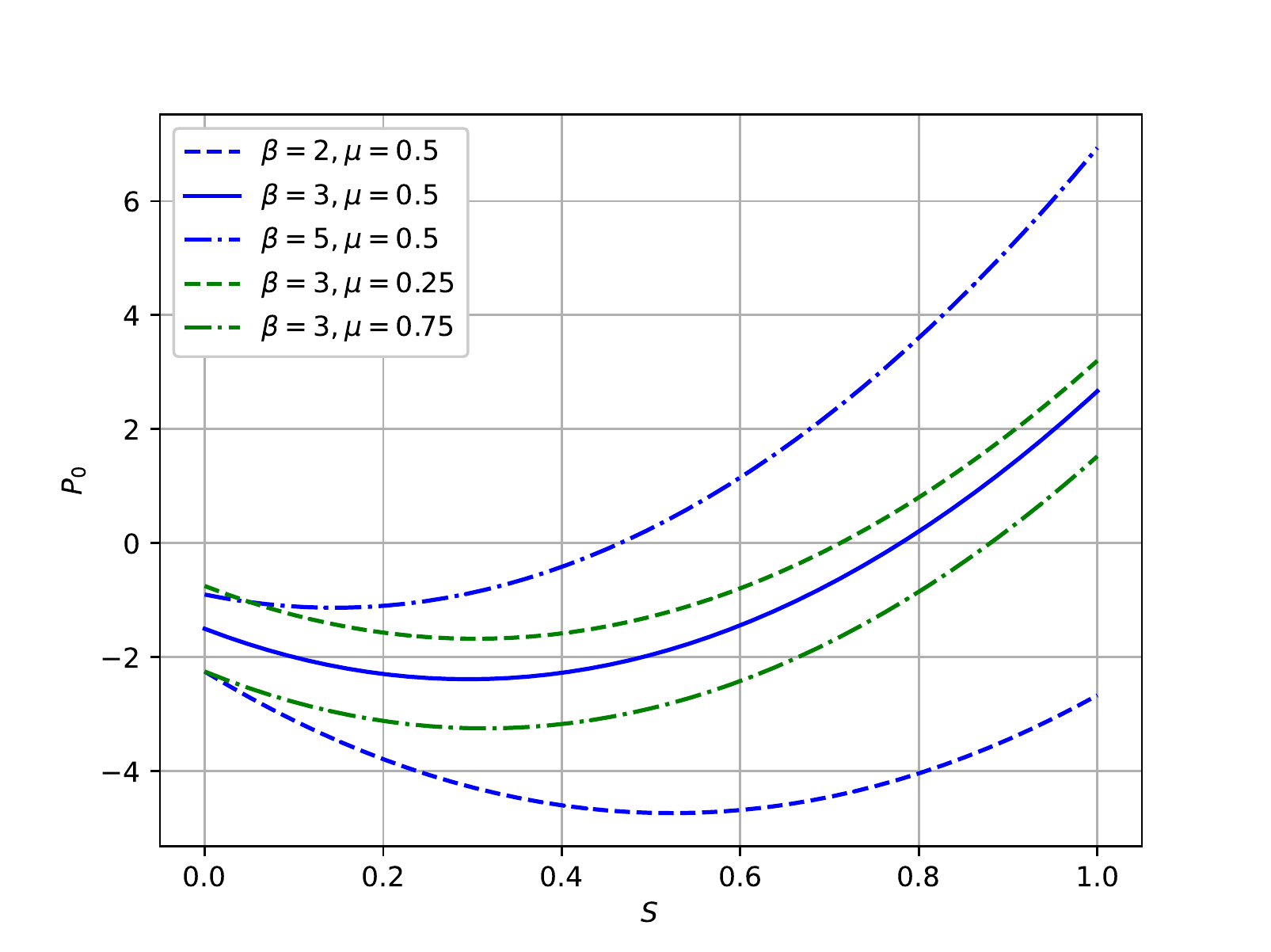}}
\caption{\label{F:1}Plot of $P_0(S)$ for $\gamma=1$ and different combinations of $\beta$ and $\mu$. Note, that for $\mu=0.5$ and $\beta=2<\tilde{\beta}_0$ (blue dashed), there exists \emph{no} equilibrium $P_0(S_0^\ast)=0$ in the interval $0\le S_0^\ast\le 1$.}
\end{figure}

Expanding by $S(S+3)$ yields the quadratic polynomial
\begin{equation*}
	P_0(S) = S(S+3)\cdot G_0(S) 
		= \lb 12 + C_0\rb S^2 + \lb 3C_0 - \frac{3\mu}{\beta}\rb S - \frac{9\mu}{\beta}\;.
\end{equation*}
Its roots determine the endemic equilibria $S_0^\ast$ of the local model. We note, that $P_0(0)=-9\mu/\beta<0$ and $P_0(1)=12+4C_0-12\mu/\beta$. For $12+C_0>0$, i.e.~$\beta>\bar{\beta}=\frac{4\gamma+5\mu}{11\gamma+15\mu}(\gamma+\mu)$ there exists a unique root $S_0^\ast>0$ given by 
\begin{equation}
\label{E:S0stern}
	S_0^\ast = \frac{3}{2(12+C_0)} \lsb \frac{\mu}{\beta}-C_0 
		+ \sqrt{\lb C_0-\frac{\mu}{\beta}\rb^2+\frac{4\mu}{\beta}(12+C_0)}\rsb\;.
\end{equation}
As can be seen in Fig.~\ref{F:1}, for fixed $\mu$ (blue curves) the transmission parameter $\beta$ needs to be large enough to obtain a root $S_0^\ast$ in the relevant interval $[0,1]$. For $\mu=0.5$ and $\beta=2$ (blue dashed curve), there exists no equilibrium $S_0^\ast<1$, while for $\beta=3$ (blue solid) the endemic equilibrium $S_0^\ast\approx 0.8$ exists. The critical value $\tilde{\beta}_0$ such that the endemic equilibrium exists,  is determined by the condition $P_0(S=1)=0$ for $\beta=\tilde{\beta}_0$, i.e.
\begin{equation*}
	0 = 2 + \frac{4\mu}{\gamma+\mu} - \frac{4\gamma+8\mu}{\tilde{\beta}_0} \quad \iff \quad
	\tilde{\beta}_0 = 2 \frac{(\gamma+2\mu)(\gamma+\mu)}{\gamma+3\mu}\,.
\end{equation*}
For $\beta>\tilde{\beta}_0$ we obtain $P_0(1)>0$ and hence a root $S_0^\ast$ in the interval $(0,1)$. For $\gamma=1$ and $\mu=1/2$, this critical value equals $\tilde{\beta}_0=12/5=2.4$.

\begin{figure}[htb]
\includegraphics[width=.475\textwidth]{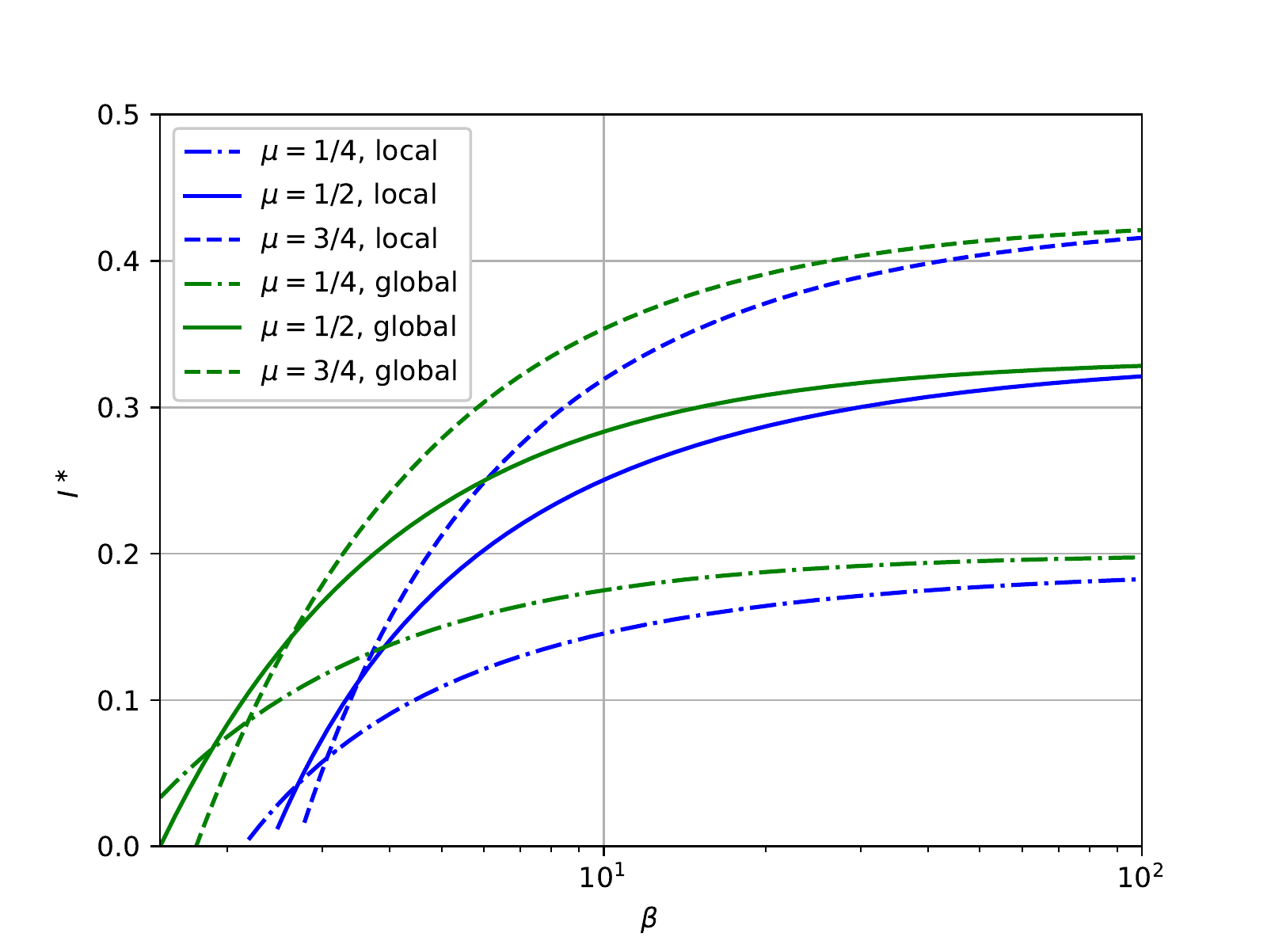} \hfill
\includegraphics[width=.475\textwidth]{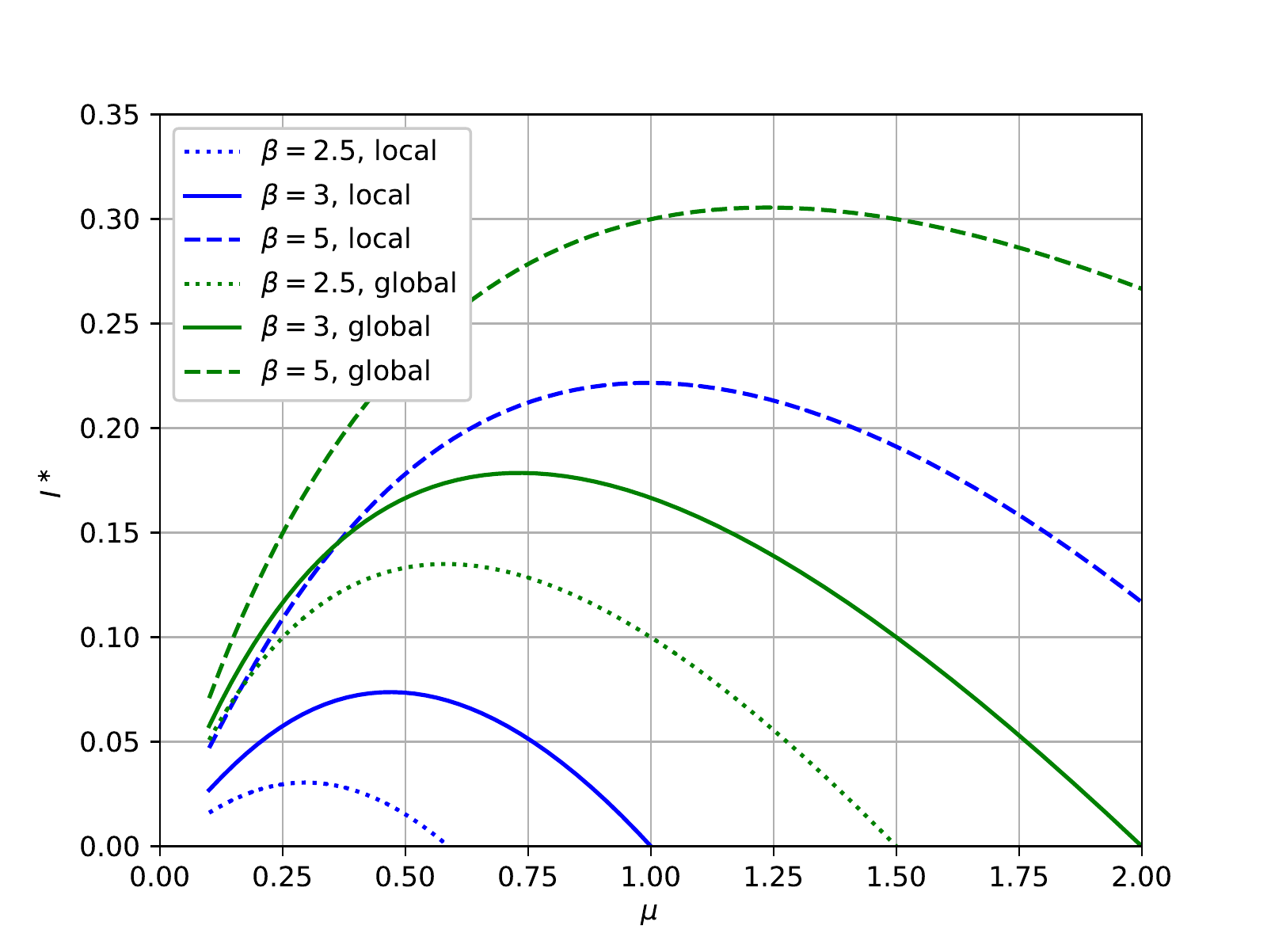}
\caption{\label{F:2} Left: Plot of $I^\ast$ vs.~$\beta$ for $\mu=1/4$, $1/2$, $3/4$. The blue curves correspond to the local model $p=0$, the green curves are for the global model $p=1$.\\
Right: Plot of $I^\ast$ vs.~$\mu$ for $\beta=2.5$, $3$, $5$.}
\end{figure}

The equilibrium $S_0^\ast$, see Eqn.~\ref{E:S0stern}, gives rise to the corresponding equilibrium value $I_0^\ast = \frac{\mu}{\gamma+\mu}(1-S_0^\ast)$ for the infected compartment. In Fig.~\ref{F:2} we compare the local equilibrium $I_0^\ast$ with its global counterpart $I_1^\ast=\frac{\mu}{\gamma+\mu}-\frac{\mu}{\beta} \xrightarrow{\beta\to\infty}\frac{\mu}{\gamma+\mu}$. 

To analyze the limiting behavior of $I_0^\ast$ for $\beta\to\infty$, we consider the limit of the explicit solution formula~\eqref{E:S0stern} for $\beta\to \infty$ in the two cases
\begin{align*}
	S_0^\ast & \xrightarrow{\beta\to\infty} \begin{cases}
		3 \frac{\gamma-3\mu}{11\gamma+15\mu} & \text{for } \gamma>3\mu \\
		0 & \text{for } \gamma\le 3\mu
	\end{cases}
\intertext{and hence}
	I_0^\ast & \xrightarrow{\beta\to\infty} \begin{cases}
		\frac{8\mu(\gamma+3\mu)}{(\gamma+\mu)(11\gamma+15\mu)} & \text{for } \gamma>3\mu \\
		\frac{\mu}{\gamma+\mu} & \text{for } \gamma\le 3\mu
	\end{cases}\;.
\end{align*}

\begin{figure}[htb]
\centerline{\includegraphics[width=.67\textwidth]{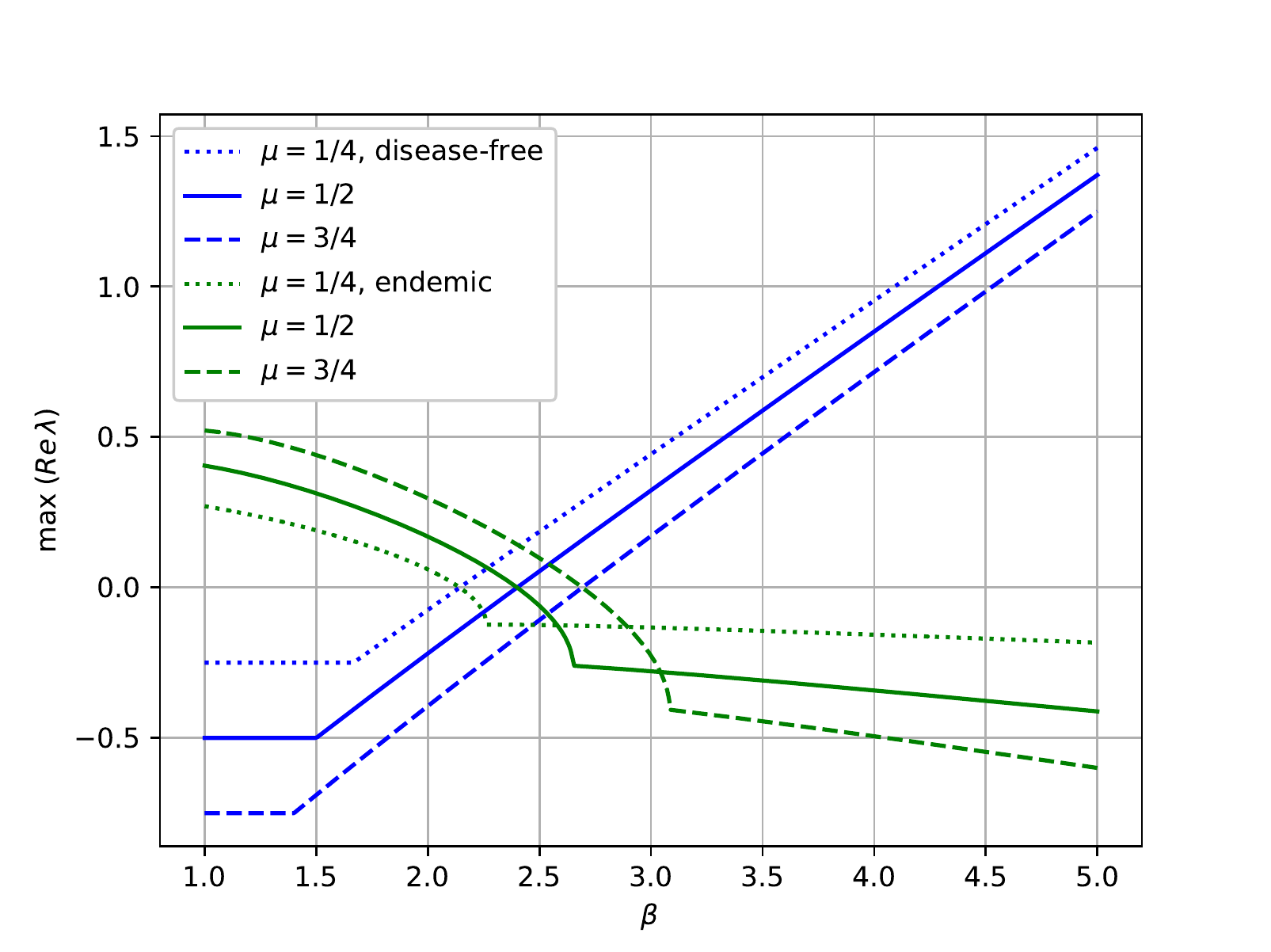}}
\caption{\label{F:EVal_J0} Maximal real part of the eigenvalues of the Jacobian $J_0$ vs.~$\beta$ for different values of $\mu$. The blue curves correspond to the disease--free equilibrium and the green ones are for the endemic equilibrium. For $\beta>\tilde{\beta}$, the disease--free equilibrium gets unstable and the endemic equilibrium turns stable.}
\end{figure}

Summing up, there exist two equilibria in the local case:
\begin{enumerate}
\item the trivial disease--free equilibrium $(S_0^0, I_0^0, X_0^0, Y_0^0)=(1,0,2,0)$ and
\item for $\beta>\tilde{\beta}_0$ the endemic equilibrium
\begin{equation*}
	(S_0^\ast, I_0^\ast, X_0^\ast, Y_0^\ast) = \lb S_0^\ast, \frac{\mu}{\gamma+\mu} (1-S_0^\ast), 
	\frac{8 S_0^\ast{}^2}{S_0^\ast+3}, \frac{4\mu}{\beta}(1-S_0^\ast)\rb\;.
\end{equation*}
\end{enumerate}
To analyze their stability, consider the Jacobian $J_{p=0}$, see~Eqn.~\eqref{E:J_p} at the disease--free equilibrium
\begin{align*}
	J_0^0 &= \lpm -\mu & 0 & 0 & -\beta/4 \\ 0 & -(\gamma+\mu) & 0 & \beta/4 \\
		4\mu & 0 & -2\mu & -3\beta/4 \\ 0 & 4\mu & 0 & -(\gamma+2\mu)+\beta/2 \rpm\;.
\end{align*}
The eigenvalues of $J_0^0$ are given by
\begin{equation*}
	\lambda_1 = -\mu, \quad \lambda_2=-2\mu \quad \text{and}\quad
	\lambda_{3,4} = \frac{\beta}{4}-\gamma-\frac{3\mu}{2} \pm \frac{1}{4} \sqrt{\beta^2+12\beta\mu+4\mu^2}\;.
\end{equation*}
For $\beta>\tilde{\beta}_0=2 \frac{(\gamma+2\mu)(\gamma+\mu)}{\gamma+3\mu}$, the maximal real part gets positive and the disease--free equilibrium gets unstable. This can be seen in Fig.~\ref{F:EVal_J0}, where we plotted the maximal real part of the eigenvalues at the disease--free equilibrium (blue curves) and at the endemic equilibrium (green curves).

\subsection{General Case $0<p<1$}

Next we consider the equilibria for the general case $0<p<1$. Recall, that in this situation we have
\begin{align*}
	I(S) &= \mu(1-S)\cdot \frac{1}{\gamma+\mu} \\
	Y(S) &= \mu(1-S)\cdot \frac{4}{q\beta} \lb 1- \frac{p\beta}{\gamma+\mu}S\rb 
		= \frac{4\mu}{\gamma+\mu} \frac{p}{q} S^2 + \dots
\end{align*}
Plugging these expressions into~\eqref{E:Fp} and factoring out the common term $\mu(1-S)$ we obtain
\begin{align*}
	0 &\stackrel{!}{=} F_p(S) 
		= 4\mu I - 2\mu Y + p\beta I (2X-Y) + \frac{3}{16} q\beta \frac{Y}{S}(2X-Y) - Y\lb \frac{q\beta}{4}+\gamma\rb \\
		&= \mu(1-S) \lcb \frac{4\mu}{\gamma+\mu} - \frac{8\mu}{q\beta} \lb 1- \frac{p\beta}{\gamma+\mu}S\rb
			+ \frac{p\beta}{\gamma+\mu}(2X-Y)\right. \\ 
		& \qquad \qquad \qquad \left. 
			+ \frac{3}{4S}\lb 1- \frac{p\beta}{\gamma+\mu}S\rb (2X-Y)
			- \frac{4}{q\beta}\lb 1- \frac{p\beta}{\gamma+\mu}S\rb \lb \frac{q\beta}{4}+\gamma\rb \rcb\\
		&=: \mu(1-S) \cdot G_p(S)
\end{align*}
where
\begin{align}
	G_p(S) &= \frac{4\mu}{\gamma+\mu} - \lb 1+\frac{8\mu+4\gamma}{q\beta}\rb\lb 1- \frac{p\beta}{\gamma+\mu}S\rb
	+ \lb \frac{1}{4}\frac{p\beta}{\gamma+\mu} + \frac{3}{4S}\rb (2X-Y)\;.
\end{align}
Multiplying $G_p$ with $S\cdot Q(S):=S\cdot \lb S+3+\frac{p\beta}{\gamma+\mu}S(1-S) \rb>0$ for $S>0$ to get rid of $S$ in the denominator of $X$ yields the following polynomial in $S$
\begin{multline*}
	P_p(S) := \lsb \frac{4\mu}{\gamma+\mu} - \lb 1+\frac{8\mu+4\gamma}{q\beta}\rb\lb 1- \frac{p\beta}{\gamma+\mu}S\rb\rsb\cdot S\cdot Q(S) \\
	 + \lsb \frac{p\beta}{4(\gamma+\mu)}S +\frac{3}{4} \rsb  \cdot \lb 16S^2-Y\cdot Q(S)\rb\;.
\end{multline*}
Since $Q$ and $Y$ are quadratic in $S$, we see, that $P_p\in \calP_5(S)$, i.e.~a polynomial of degree $5$. Hence we may expect up to $5$ possible roots. To determine whether some of them are in the relevant interval $0<S<1$ we note that
\begin{align*}
	P_p(S=0) &= - \frac{9\mu}{q\beta} < 0
\intertext{and}
	P_p(S=1) &= 8 + \frac{16\mu+8p\beta}{\gamma+\mu} + 4 \frac{8\mu+4\gamma}{q\beta} \lb \frac{p\beta}{\gamma+\mu}-1\rb\;.
\end{align*}
Recasting the terms, we can write $P_p$ as
\begin{multline}
\label{E:Pp}
	P_p(S) = 4\frac{p\beta}{\gamma+\mu}S^3 + 12 S^2 \\
	+ Q(S)\cdot 
		\lsb \frac{4\mu}{\gamma+\mu}S - \lb \frac{4\mu}{q\beta}+S+\frac{4(\gamma+\mu)}{q\beta}S-\frac{1}{4}Y\rb \lb 1- \frac{p\beta}{\gamma+\mu}S\rb\rsb
\end{multline}
The leading coefficient of this polynomial is given by
\begin{equation*}
	a_5 = \frac{p\beta}{\gamma+\mu}\cdot \frac{1}{4}\cdot \frac{4\mu}{\gamma+\mu}\cdot \frac{p}{q}
				\cdot \frac{p\beta}{\gamma+\mu}
	= \frac{p^3\beta^2}{q(\gamma+\mu)^3} >0 
\end{equation*}
Hence there exists at least one positive root of $P_p$, the endemic equilibrium. Since no analytic solutions are available, we solve the polynomial $P_p$ numerically to identify the equilibira.

\begin{figure}[htb]
\includegraphics[width=.475\textwidth]{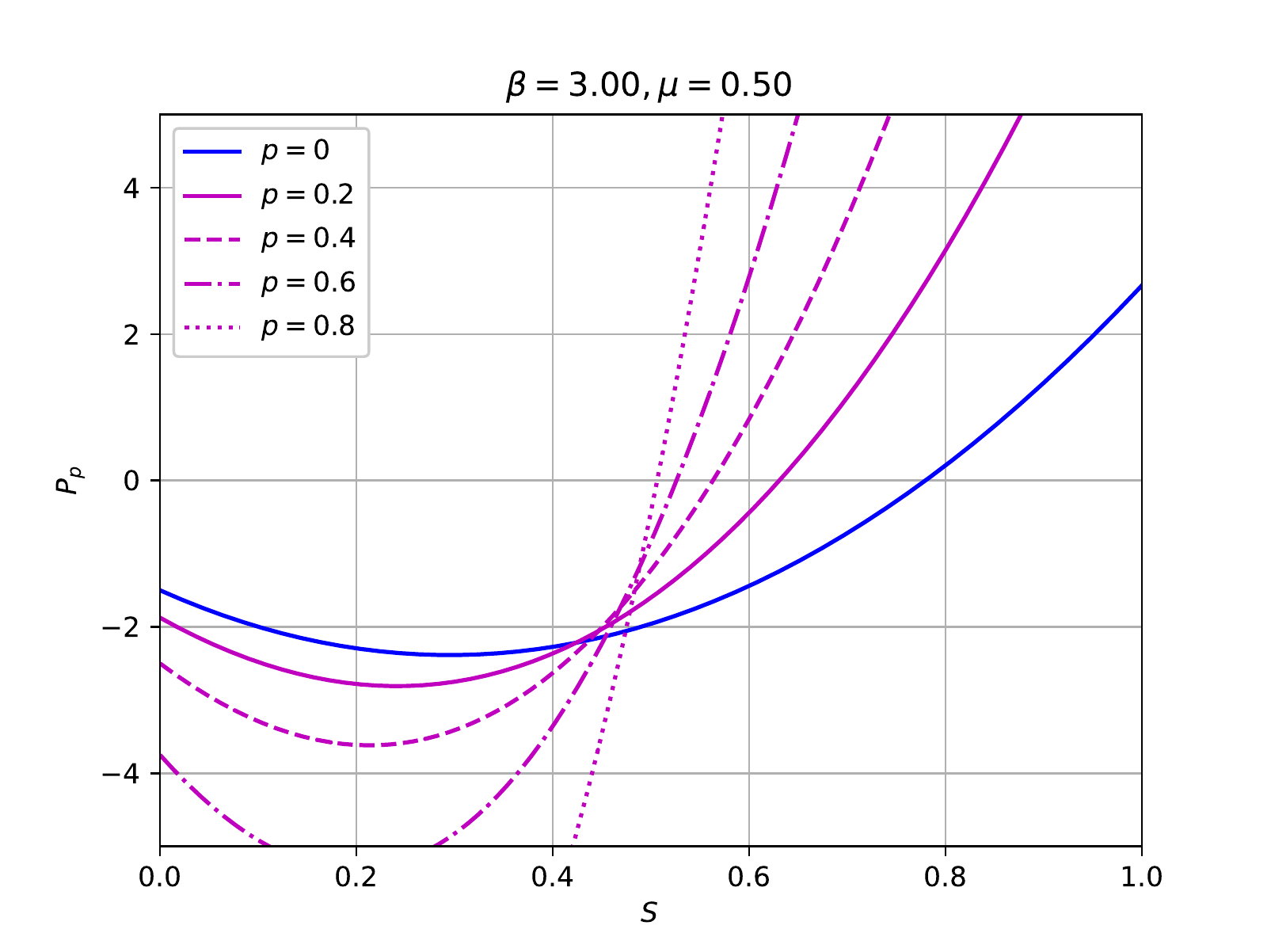} \hfill
\includegraphics[width=.475\textwidth]{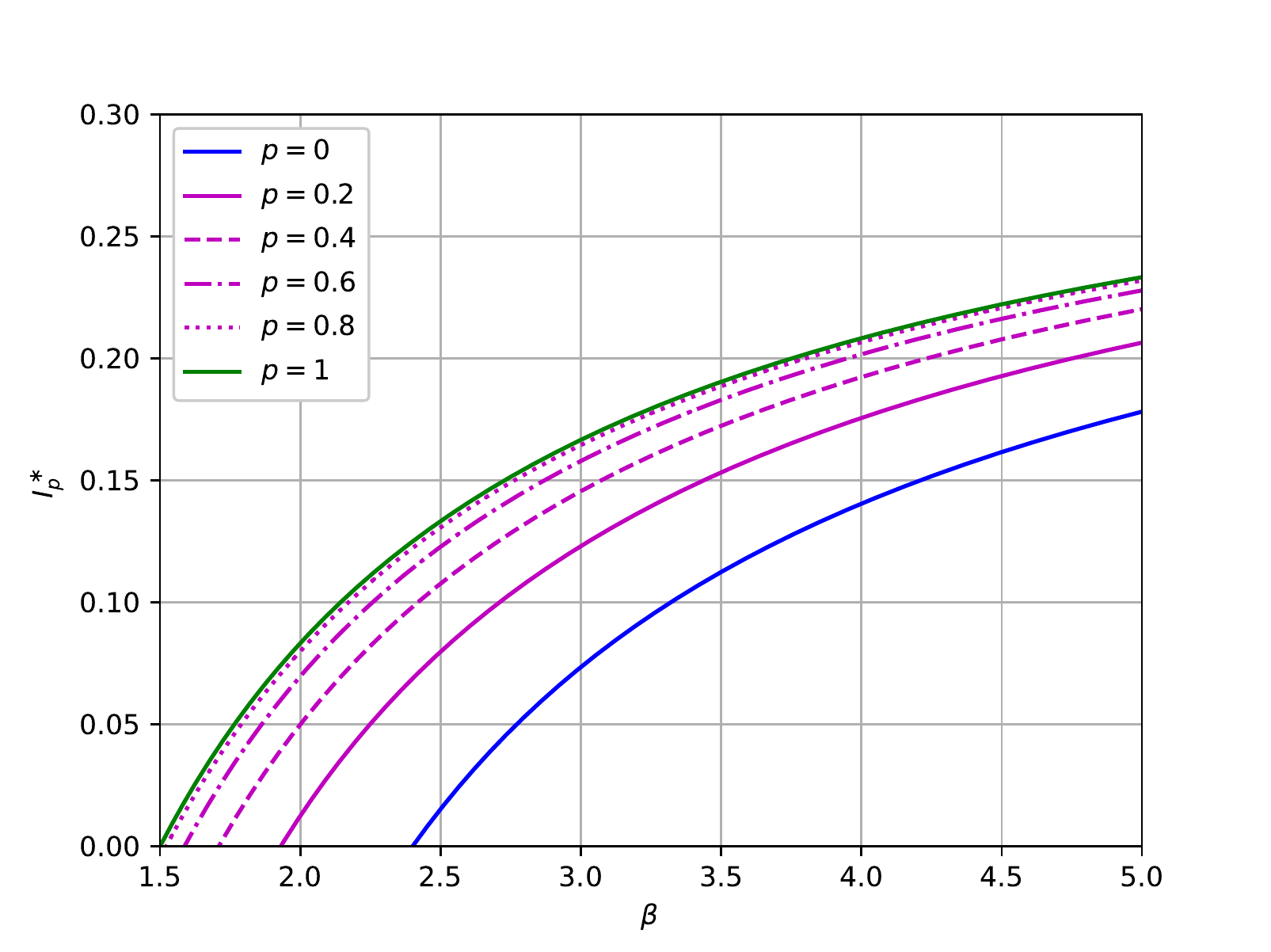}
\caption{\label{F:Pp_Ip} Left: Plot of the polynomial $P_p(S)$ for $\beta=3$, $\gamma=1$ and $\mu=1/2$ fixed. The locality parameter $p$ varies between $p=0$ (solid blue) and $p=0.8$ (dotted magenta). The zeros correspond to the endemic equilibrium.\newline Right: Endemic equilibrium $I_p^\ast$ vs.~$\beta$ for $\gamma=1$ and $\mu=1/2$ fixed and $3/2\le \beta \le 5$. Locality parameter $p$ varies between $p=0$ (solid blue) and $p=1$ (solid green).}
\end{figure}

In Fig.~\ref{F:Pp_Ip} we have plotted the numerical results. The equilibrium $I_p^\ast$ given in Fig.~\ref{F:Pp_Ip}(right) for $\beta=3$ corresponds to the stationary state visible in the transient solution in Fig~\ref{F:Ip_vs_t}. 

\begin{figure}[htb]
\centerline{\includegraphics[width=0.67\textwidth]{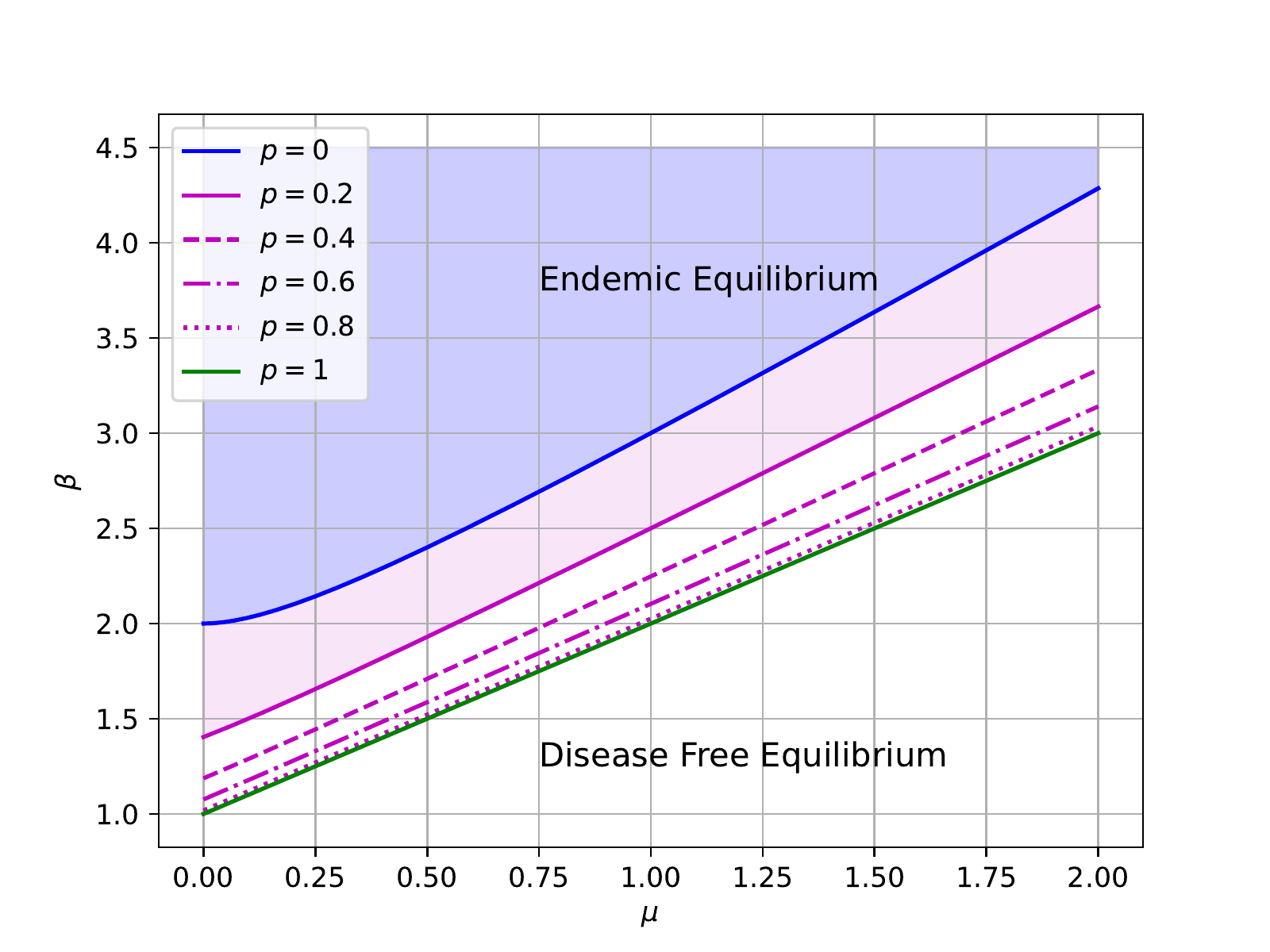}}
\caption{\label{F:tilde_beta} Critical value $\tilde{\beta}_p$ vs. $\mu$ for different values of the locality parameter $p$, recovery rate $\gamma=1$ fixed. For $\beta>\tilde{\beta}_p$ (e.g. pale blue area for $p=0$), the endemic equilibrium exists and is stable. For smaller values of $\beta$, there exists only the disease--free equilibrium.}
\end{figure}

To ensure the existence of the endemic equilibrium $I_p^\ast$, the following condition $P_p(S=1)>0$ is necessary. Solving 
\begin{gather*}
	P_p(S=1) = \frac{8}{q\beta(\gamma+\mu)} \lsb \beta(3\mu+\gamma+p\beta) + p\beta(\mu+\gamma-p\beta) - (4\mu+2\gamma)(\gamma+\mu)\rsb > 0
\intertext{for $\beta$ while keeping $\gamma,\mu>0$ fixed yields}
	\beta^2 (p-p^2) + \beta ((3+p)\mu+(1+p)\gamma)-2(2\mu+\gamma)(\gamma+\mu)>0\;.
\end{gather*}
Since $p-p^2>0$, there exists a critical value $\tilde{\beta}_p>0$, such that $P_p(S=1)>0$ for $\beta>\tilde{\beta}_p$. This critical value is given by
\begin{multline}
\label{E:tilde_beta_p}
	\tilde{\beta}_p := \frac{1}{2(p-p^2)} \lsb
		\sqrt{((3+p)\mu+(1+p)\gamma)^2+8(2\mu+\gamma)(\gamma+\mu)(p-p^2)} \right.\\
		\left. -((3+p)\mu+(1+p)\gamma)\rsb \;.
\end{multline}
In the two extremal cases $p=0$ and $p=1$ we reobtain our previous findings for the critical value
\begin{align*}
	\tilde{\beta}_0 &:= 2 \frac{\gamma+2\mu}{\gamma+3\mu}(\gamma+\mu)\;, \\
	\tilde{\beta}_1 &:= \gamma+\mu\;.
\end{align*}
In Figure~\ref{F:tilde_beta} we have depicted the regions, where the endemic equilibrium exists. For fixed $\mu$ this region is determined by $\beta>\tilde{\beta}_p$. For smaller values of $\beta$, the only equilibrium is the disease--free equilibrium.

Inspecting the Jacobian
\begin{equation*}
	J_p^0 = \lpm -\mu & -p\beta & 0 & -q\beta/4 \\ 
		0 & p\beta-(\gamma+\mu) & 0 & q\beta/4 \\
		4\mu & -4p\beta & -2\mu & -3q\beta/4 \\ 
		0 & 4\mu+4p\beta & 0 & -(\gamma+2\mu)+q\beta/2 \rpm
\end{equation*}
at the disease--free equilibrium yields the following eigenvalues $\lambda_1=-\mu$, $\lambda_2=-2\mu$ and
\begin{equation*}
	\lambda_{3,4} = \frac{1}{4} \lsb (1+p)\beta-4\gamma-6\mu \pm \sqrt{((1+p)\beta+2\mu)^2+8pq\beta^2+8q\mu\beta}\rsb\;.
\end{equation*}
For $\beta>\tilde{\beta}_p$, see Eqn.~\eqref{E:tilde_beta_p}, the maximal real part of $\lambda_3$ gets positive. Again, as in the purely local case $p=0$, as soon as the endemic equilibrium appears, the disease--free equilibrium gets unstable.

\section{Basic Reproduction Number}
\label{S:R0}

To compute the basic reproduction number we follow the next generation matrix approach, see~\cite{DriWat02, Mar15}. We subdivide the system~\eqref{E:Pairmodel} into the infected compartments $\xi=(I,Y)$ and the non--infected ones $\eta=(S,X)$. The equations for the infected compartments are written as
\begin{align*}
	\xi ' &= \calF(\xi,\eta) - \calV(\xi,\eta)
\intertext{where} 
	\calF_p(\xi,\eta) &= \lpm p\beta\xi_1 \eta_1 + q\beta\xi_2/4 \\
		2p\beta \xi_1\eta_2 + \tfrac{3}{8} q\beta \xi_2 \eta_2/\eta_1 \rpm
\intertext{describes the new infections and}
	\calV_p(\xi,\eta) &= \lpm (\gamma+\mu)\xi_1 \\
		p\beta\xi_1\xi_2 + \frac{q\beta}{4}\xi_2\lb \frac{3\xi_2}{4\eta_1}+1\rb
		+(\gamma+\mu)\xi_2 - \mu(4\xi_1-\xi_2) \rpm
\end{align*}
describes other transitions. The disease free equilibrium is given by $\xi^0=(0,0)$ and $\eta^0=(1,2)$. Introducing the Jacobians $F_p=\frac{\pder}{\pder\xi} \calF_p(0, \eta^0)$ and $V_p=\frac{\pder}{\pder\xi} \calV_p(0, \eta^0)$, the basic reproduction number is given as the spectral radius of the next generation matrix $M_p=F_p V_p^{-1}$, i.e.
\begin{equation*}
	\calR_{0,p} = \rho(M_p) = \max \lcb \abs{\lambda}:\ \text{$\lambda$ eigenvalue of $M_p$}\rcb\;. 
\end{equation*}
In our case
\begin{align*}
	F_p &= \lpm p\beta & \frac{1}{4}q\beta \\[.5ex]
		4p\beta & \frac{3}{4}q\beta \rpm\;, \quad
	V_p = \lpm \gamma+\mu & 0 \\
		-4\mu & \frac{q\beta}{4}+\gamma+2\mu \rpm
\intertext{and}
	M_p &= \lpm p\beta & \tfrac{1}{4}q\beta \\[.5ex]
		4p\beta & \tfrac{3}{4}q\beta \rpm \cdot
		\lpm \frac{1}{\gamma+\mu} & 0 \\[.5ex]
		\frac{4\mu}{(\gamma+\mu)(\gamma+2\mu+q\beta/4)} & 
		\frac{1}{\gamma+2\mu+q\beta/4} \rpm \\
		&= \lpm pk_1+ q\mu k_1/k_2 & q\beta/(4 k_2) \\
			4pk_1 + 3q\mu k_1/k_2 & 3q\beta/(4 k_2) \rpm
\end{align*}
where $k_1=\beta/(\gamma+\mu)$ and $k_2=\gamma+2\mu+q\beta/4$.
Hence 
\begin{equation*}
	\calR_{0,p} = \frac{1}{8k_2}\lsb 4pk_1 k_2 + q (3\beta+ 4\mu k_1) 
	+ \sqrt{(4pk_1 k_2 + q(3\beta+4\mu k_1))^2+16pq\beta k_1 k_2}\rsb
\end{equation*}
In the global case $p=1$, $q=0$, this reduces to the well--known basic reproductive number for the classical SIR--system
\begin{equation*}
	\calR_{0,p=1} = k_1 = \frac{\beta}{\gamma+\mu}
\end{equation*}
and for the local case $p=0$ and $q=1$ we get
\begin{equation*}
	\calR_{0,p=0} = \frac{3\beta+4\mu k_1}{4k_2}
		= \frac{\beta}{\gamma+\mu}\cdot \frac{3\gamma + 7\mu}{4\gamma+8\mu+\beta}
		= \calR_{0,1}\cdot \frac{3\gamma + 7\mu}{4\gamma+8\mu+\beta}
\end{equation*}
It is obvious, that $\calR_{0,p=0}<\calR_{0,p=1}$. The local transmission via pair interactions is slower than the global transmission. This is confirmed by stochastic simulations, see~\cite{MalFab16, WreBes21}.

\begin{figure}[htb]
\centerline{\includegraphics[width=0.67\textwidth]{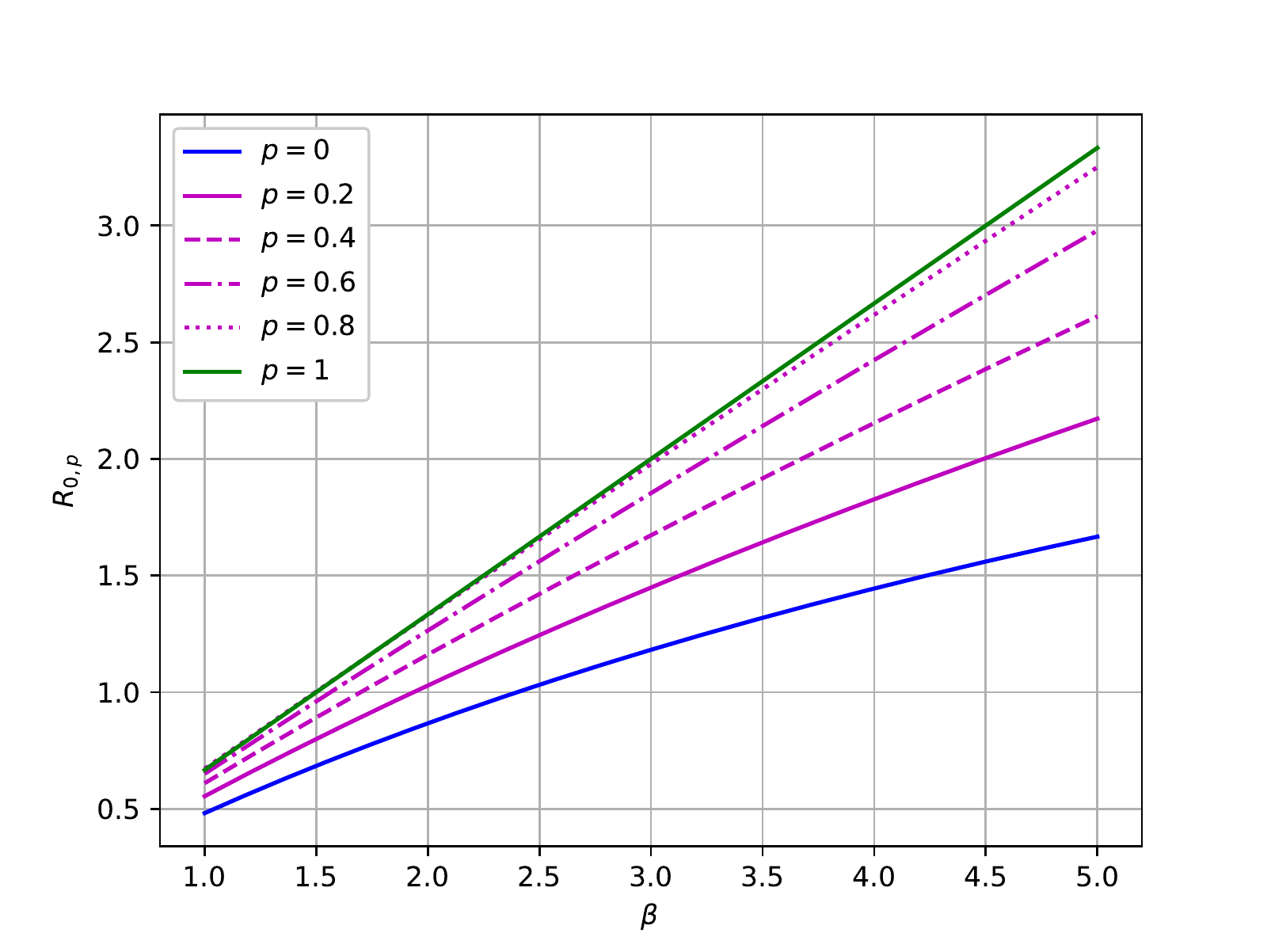}}
\caption{\label{F:R0_beta} Basic reproduction number $\calR_{0,p}$ vs.~$\beta$ for $\gamma=1$ and $\mu=1/2$ fixed. The locality parameter $p$ varies between $p=0$ (blue) and $p=1$ (green).}
\end{figure}

The critical value $\calR_{0,p=0}=1$ for the local model corresponds to
\begin{align*}
	\beta = \tilde{\beta}_0 &= 2\frac{\gamma+2\mu}{\gamma+3\mu}(\gamma+\mu)\;.
\end{align*}
In the general case $p\in (0,1)$ we get
\begin{multline}
	\calR_{0,p} = \calR_{0,1} \cdot \lsb \frac12p 
		+ q\frac{3\gamma+7\mu}{8\gamma+16\mu+2q\beta} \right. \\
		\left. + \sqrt{\lb \frac12 p+q\frac{3\gamma+7\mu}{8\gamma+16\mu+2q\beta}\rb^2 
		+ pq\frac{\gamma+\mu}{4\gamma+8\mu+q\beta}}\rsb\;.
\end{multline}
In Fig.~\ref{F:R0_beta} we show the variation of the basic reproduction number with varying $\beta$ and $p$.

\section{Transient Solutions}
\label{S:transient}

\begin{figure}[htb]
\centerline{\includegraphics[width=.67\textwidth]{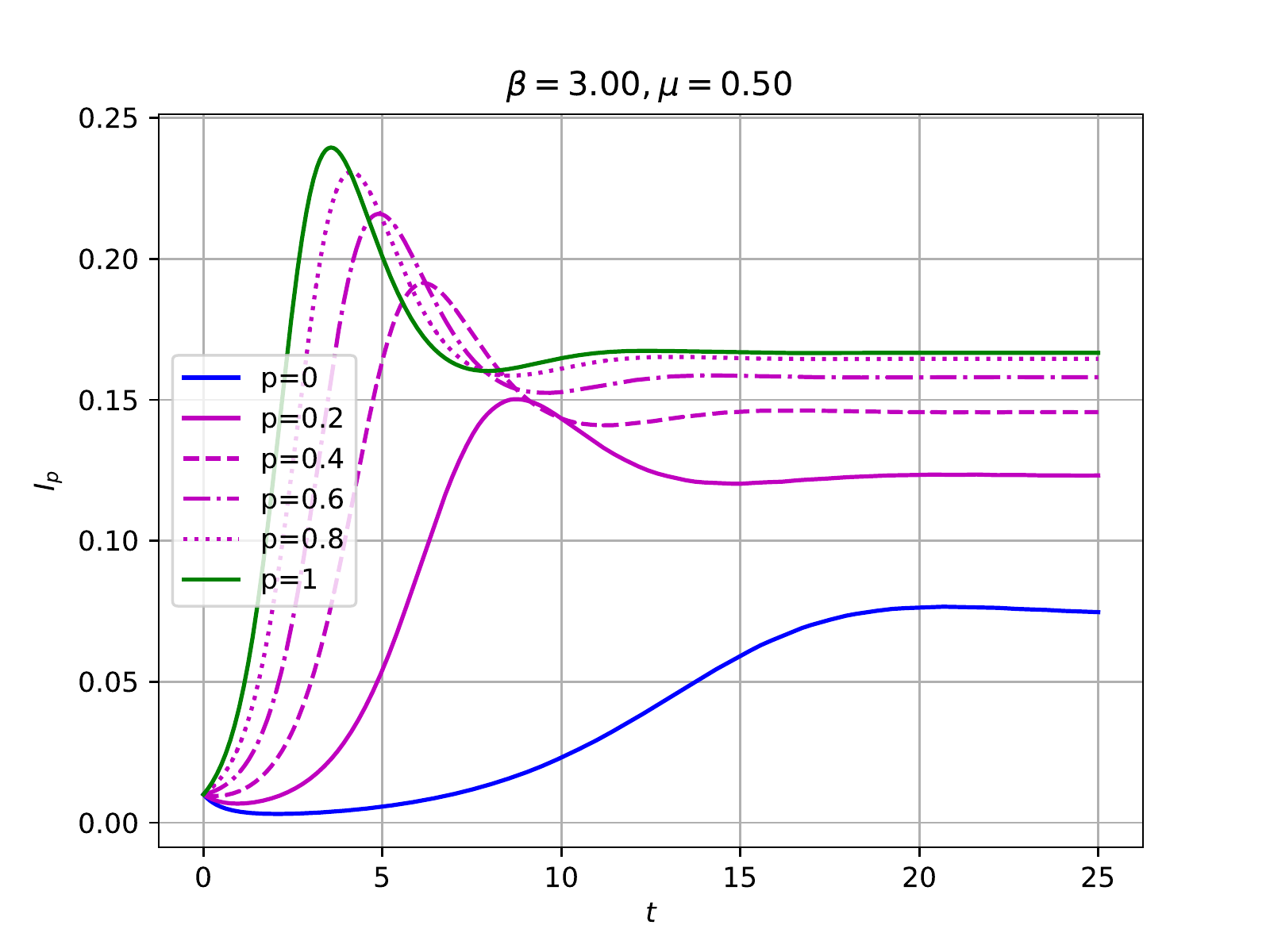}}
\caption{\label{F:Ip_vs_t} Plot of the infected compartment $I_p$ vs. time for $\beta=3$, $\gamma=1$ and $\mu=1/2$. The locality parameter $p$ varies between $p=0$ (blue) and $p=1$ (green).}
\end{figure}

We solve the ODE--system~\eqref{E:Pairmodel} numerically using a standard Runge--Kutta--Fehlberg method of order 5(4). The recovery rate $\gamma=1$ is fixed and the time interval is chosen as $t\in [0, 25]$ to ensure that the transient solutions reach the possible equilibrium. The initial condition $(S,I,X,Y)(t=0)=(0.99,0.01,0.0)$ models the scenario of  $1\%$ infections in an otherwise initially naive population. In Figure~\ref{F:Ip_vs_t} we show the transient behavior of the infected compartment $I_p$ for different values of the locality parameter $p\in [0,1]$ and $\beta=3$, $\mu=1/2$. All trajectories settle at the corresponding endemic equilibrium $I_p^\ast$ shown in Fig.~\ref{F:Pp_Ip}(right). 

\begin{figure}[htb]
\centerline{\includegraphics[width=.67\textwidth]{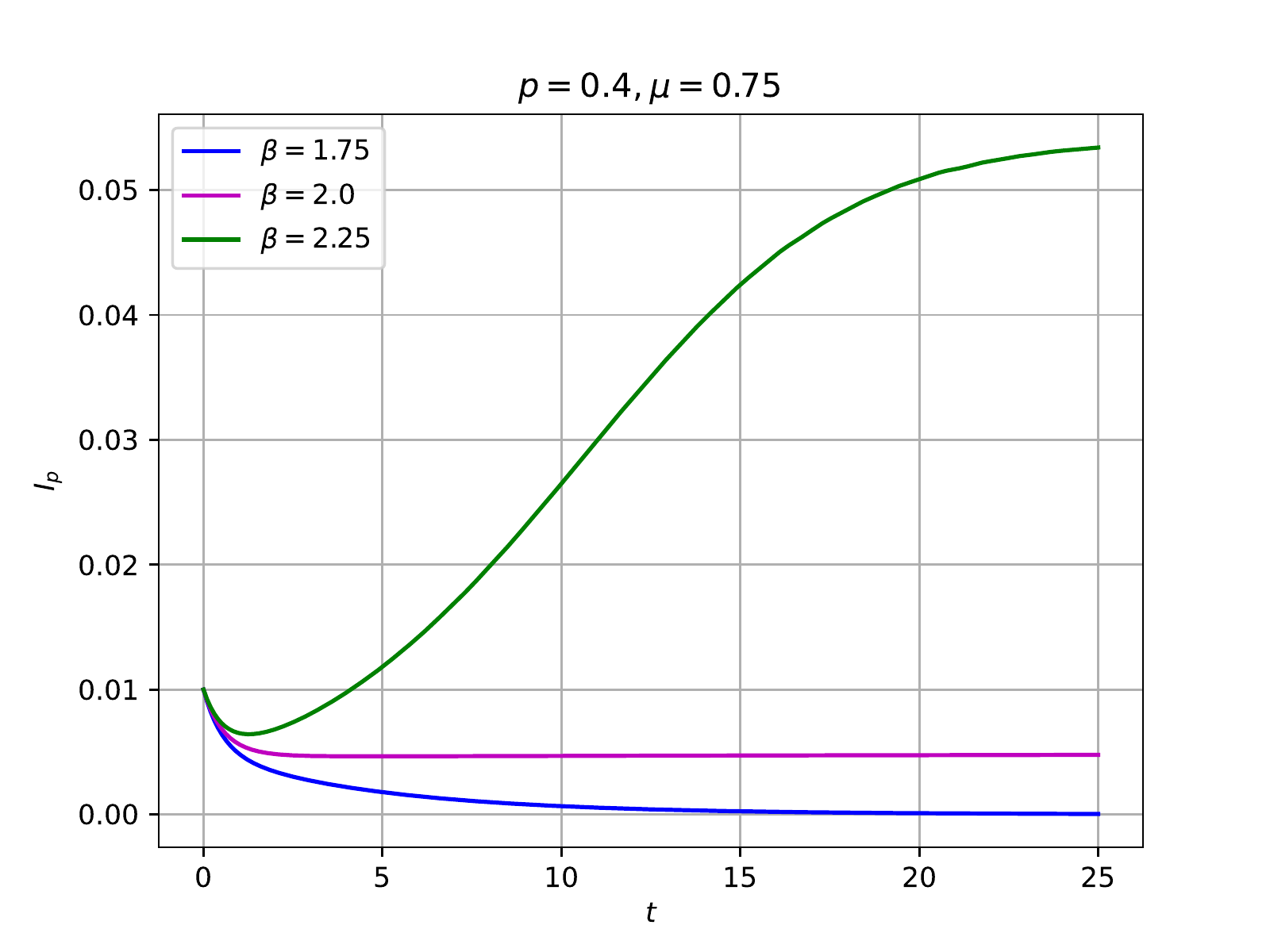}}
\caption{\label{F:Ip_vs_t_beta} Plot of the infected compartment $I_p$ vs. time for $p=0.4$, $\gamma=1$ and $\mu=0.75$. The transmission rate $\beta$ varies between the subcritical value $1.75<\tilde{\beta}$ (blue) where the disease free equilibrium is asymptotically stable and the two other cases $\beta=2$ and $2.25$ (magenta and green), where the transmission rate is above the threshold $\tilde{\beta}_{0.4}=1.978$ and hence the endemic equilibrium gets the stable one.}
\end{figure}

To visualize the convergence against the respective equilibrium, we consider in Fig.~\ref{F:Ip_vs_t_beta} the situation for $p=0.4$, $\gamma=1$ and $\mu=0.75$ fixed. According to Eqn.~\eqref{E:tilde_beta_p} or Fig.~\ref{F:tilde_beta}, for these parameter values, the critical transmission rate equals $\tilde{\beta}_{p=0.4}=1.978$. For larger values of $\beta$, the transient solution will converge to the endemic equilibrium  $I_p^\ast$ and for smaller values of $\beta$ the disease will die out since the disease--free equilibrium is asymptotically stable. The curves in Fig.~\ref{F:Ip_vs_t_beta} confirm this; the blue trajectory for $\beta=1.75<\tilde{\beta}_p$ tends to zero while the two other trajectories for $\beta=2$ (magenta) and $\beta=2.25$ (green) tend to the respective endemic equilibria. 

\begin{figure}[htb]
\centerline{\includegraphics[width=.67\textwidth]{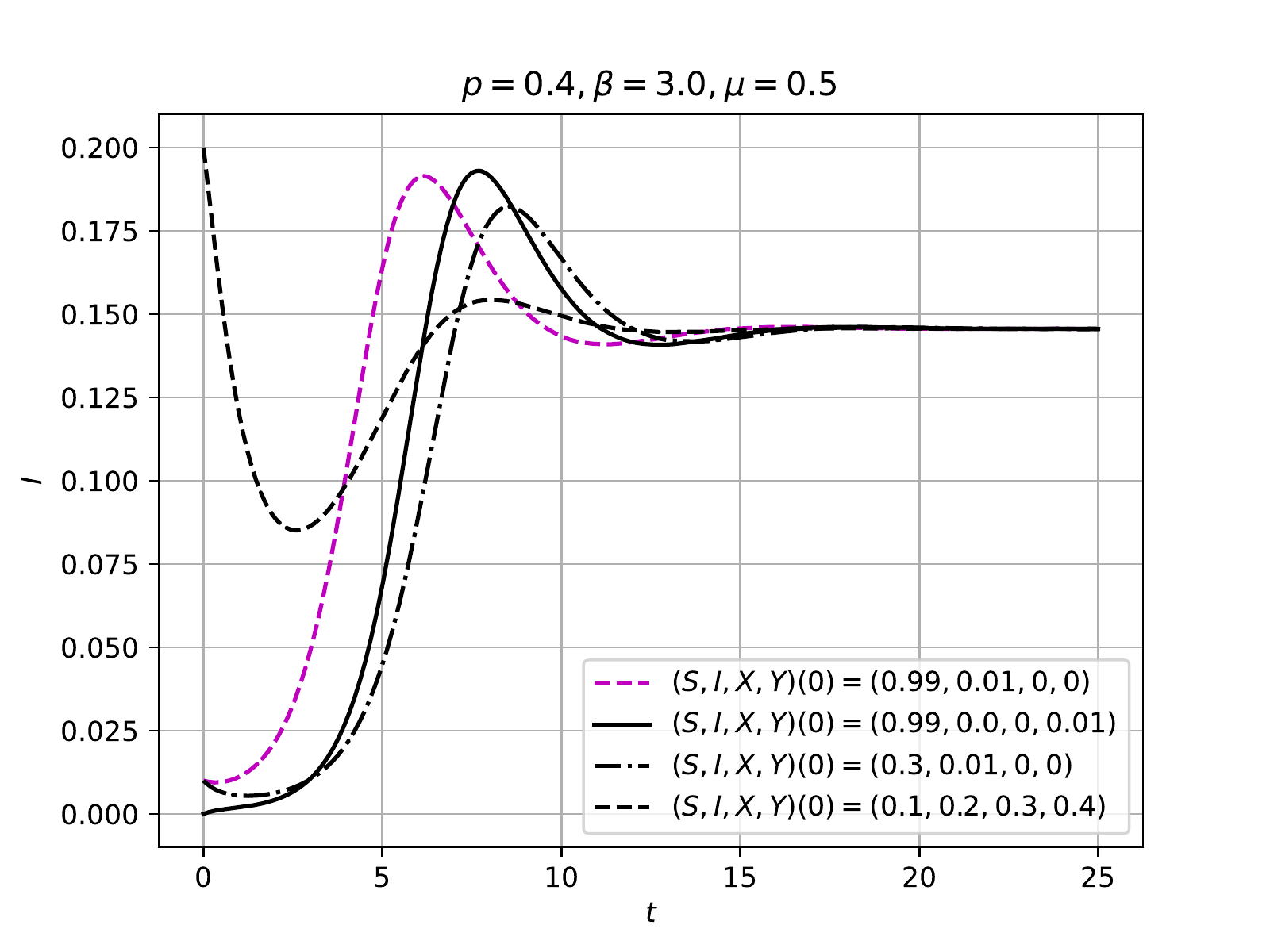}}
\caption{\label{F:I_vs_I0} Plot of the infected compartment $I$ vs.~time for varying initial conditions $(S,I,X,Y)(0)$. The locality parameter $p=0.4$ and the model parameters $\beta=3$, $\gamma=1$ and $\mu=0.5$ are fixed.}
\end{figure}

As a final check we investigated the behavior of the transient solutions for varying initial conditions. In Fig.~\ref{F:I_vs_I0} we plotted the transient solution $I$ in the mixed model $p=0.4$ for parameters $(\beta,\gamma,\mu)=(3,1,0.5)$ fixed and various initial conditions. All simulations tend to the same endemic equilibrium $I^\ast_{p=0.4}\simeq 0.146$ as to be expected. 

\section{Discussion and Conclusion}

The pair approximation model considers the infected pairs $SI$, $II$ and $IR$. The pairs $II$ and $IR$ can be considered as \emph{blocked} since no transmission can occur in that situation. The only \emph{active} pair is the combination $SI$. Hence we introduce the ratio $\rho_p$ as the ratio between active and all infection pairs. The total number of pairs including infected individuals equals $Y+2Z_{II}+Z_{IR} = 4I$ (the $II$--pair contains two infected) and we obtain 
\begin{align*}
	\rho_p &= \frac{Y}{4I}\;.
\intertext{It's equilibrium value is given by}
	\rho_p^\ast &= \frac{Y_p^\ast}{4I_p^\ast} 
		= \frac{\gamma+\mu}{q\beta} \lb 1-\frac{p\beta}{\gamma+\mu} S_p^\ast\rb
\end{align*}
for $p<1$ and $q>0$. In both extremal cases $p=1$ and $p=0$ we obtain the same equilibrium value
\begin{equation*}
	\rho_1^\ast = \rho_0^\ast = \frac{\gamma+\mu}{\beta}\;.
\end{equation*}
Computing $\rho_p$ for arbitrary $p\in (0,1)$ seems out of reach, since it requires the solution an endemic equilibrium $S_p^\ast$, the root of the quintic polynomial~\eqref{E:Pp}.
 
\begin{figure}[htb]
\centerline{\includegraphics[width=.67\textwidth]{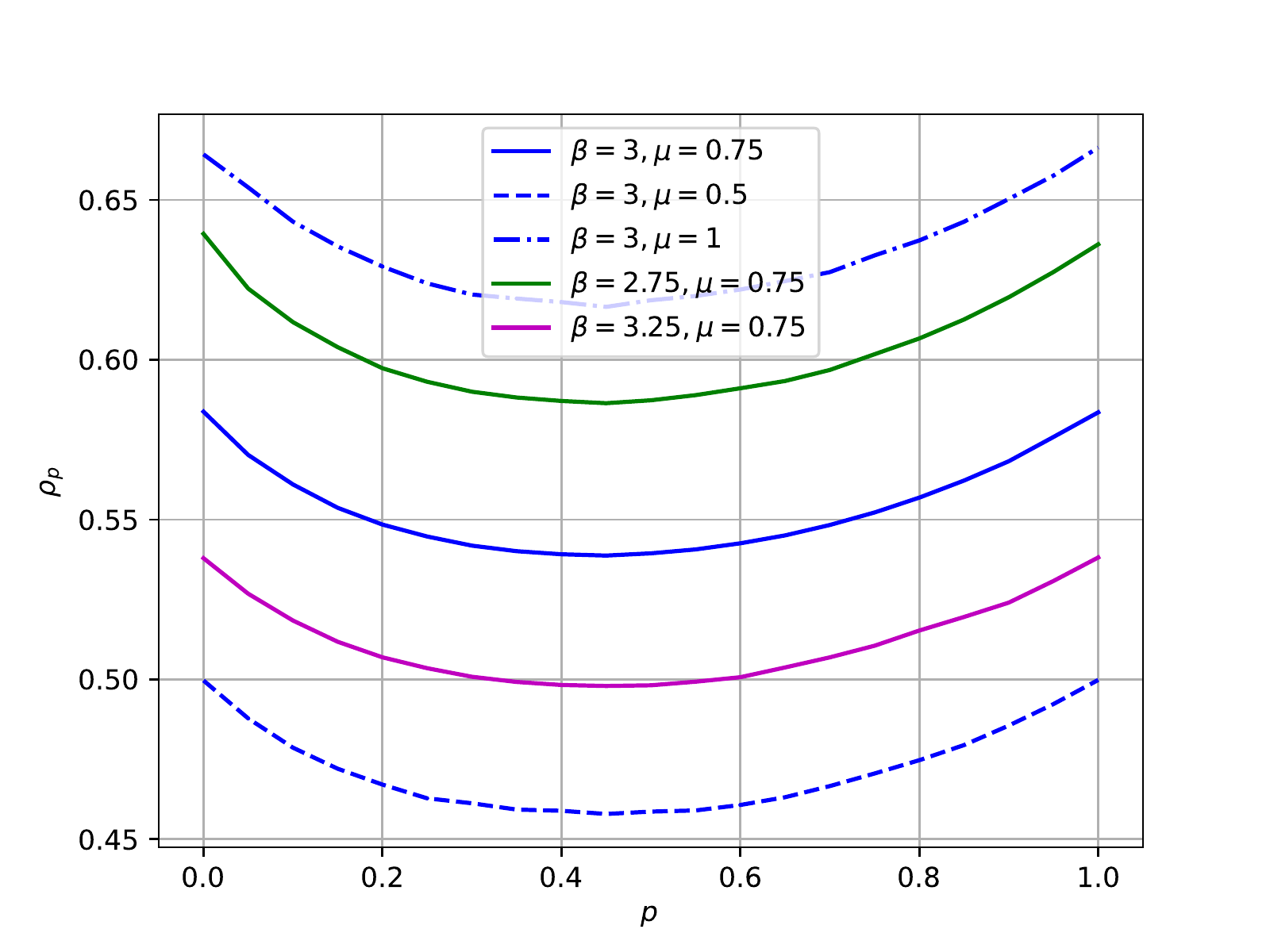}}
\caption{\label{F:rho_p} Equilibrium value of the fraction $\rho_p$ of active pairs. The model parameter $\gamma=1$ is fixed.}
\end{figure}

\begin{figure}[htb]
\centerline{\includegraphics[width=.67\textwidth]{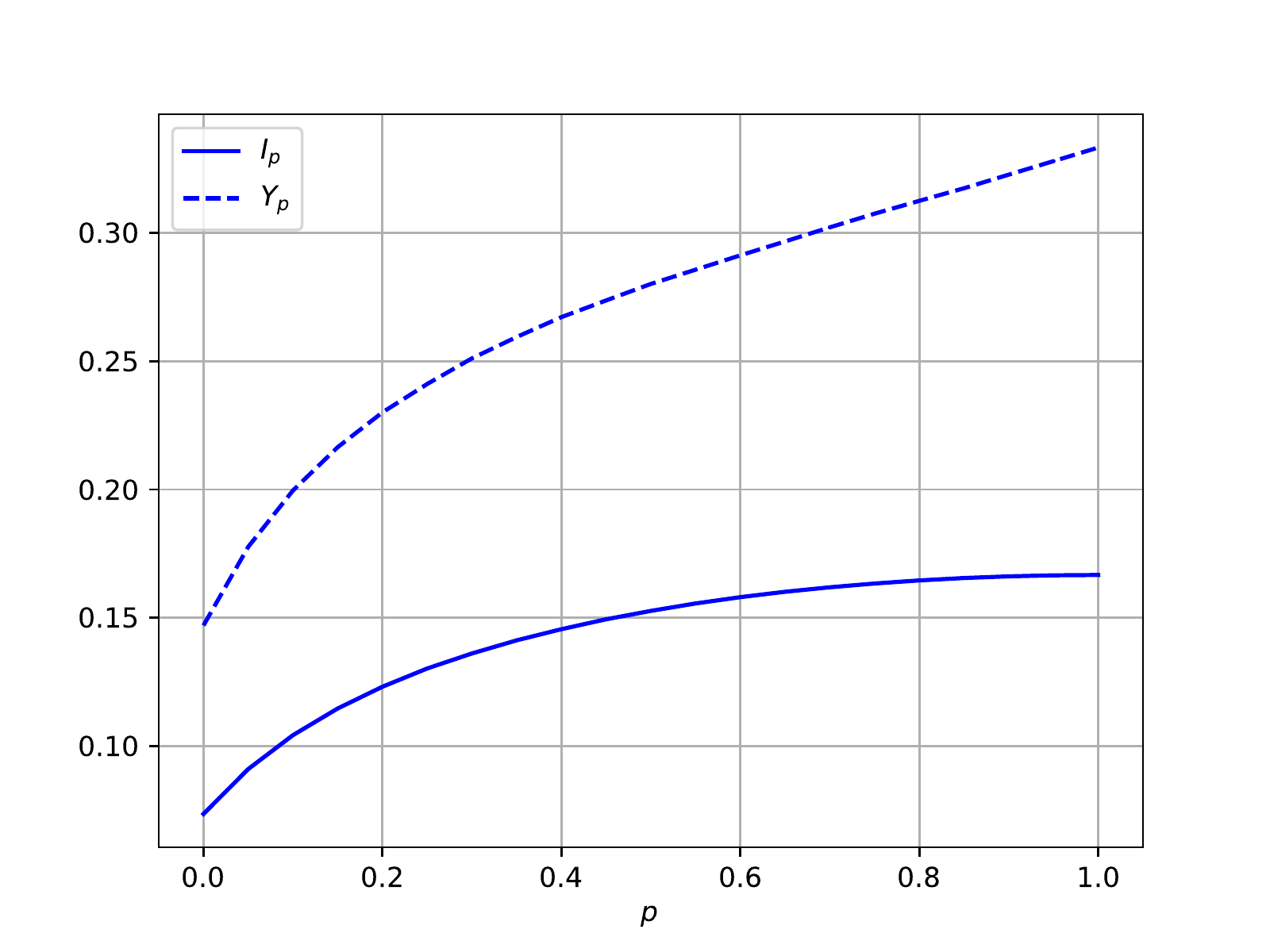}}
\caption{\label{F:IpYp_p} Endemic equilibria $I^\ast_p$ and $Y^\ast_p$ vs $p$. The model parameters $\beta=3$, $\gamma=1$ and $\mu=0.5$ are fixed.}
\end{figure}

In Fig.~\ref{F:rho_p} we show the ratio $\rho_p$ vs. the locality parameter $p$ for different disease parameters $\beta$ and $\mu$ ($\gamma=1$ is fixed). We observe, that in both extremal cases $p=0$ (local model) and $p=1$ (global model), the fraction of active pairs is maximal in each situation. For intermediate values of $p$ more pairs gets blocked and the fraction of active pairs attains a minimum. Note, that the curves are \emph{not} symmetric, i.e.~the minimum is not attained for $p=1/2$. Rewriting $\rho_p^\ast$ as
\begin{align*}
	\rho_p^\ast &= \frac{1}{1-p}\lb \frac{\gamma+\mu}{\beta} - p S_p^\ast\rb = \frac{1}{1-p}\lb S_1^\ast - p S_p^\ast\rb
\intertext{we get}
	\left.\frac{d}{dp}\rho_p^\ast\right\vert_{p=0} &= S_1^\ast-S_0^\ast\;.
\end{align*}
To show, that $S_1^\ast<S_0^\ast$ we consider
\begin{align*}
	G_0(S_1^\ast) &= \frac{12S_1^\ast}{3+S_1^\ast} - \frac{3\mu}{\beta S_1^\ast} + C_0 
		= \frac{12(\gamma+\mu)}{3\beta+\gamma+\mu} + \frac{\mu}{\gamma+\mu} - \frac{5\mu+4\gamma}{\beta} -1 \\
		&= - \frac{\beta\gamma^2+3\beta\mu^2+4\beta\gamma\mu+3\beta^2\gamma+(5\mu+4\gamma)(\gamma+\mu)^2}{\beta (3\beta+\gamma+\mu)(\gamma+\mu)}
\end{align*}
Hence $G_0(S_1^\ast)<0$ and therefore $S_1\ast<S_0^\ast$ (cf.~Fig.~\ref{F:1}) and finally $\left.\frac{d}{dp}\rho_p^\ast\right\vert_{p=0}<0$. To visualize the emergence of the minimum of $\rho_p^\ast$ graphically, we have plotted in Fig.~\ref{F:IpYp_p} the endemic equilibria $I^\ast_p$ and $Y^\ast_p$ for $p\in [0,1]$. In this graph the model parameters $\beta=3$, $\gamma=1$ and $\mu=0.5$ are fixed. As we can see, both curves decrease as  $p$ decreases from the global model $p=1$ to smaller values. However, the $SI$--pairs $Y_p^\ast$ decay faster than the total infected $I^\ast_p$; hence the ratio $\rho_p=Y_p^\ast/(4I_p^\ast)$ also decreases as $p$ gets smaller than $1$. In the neighborhood of the local situation $p=0$ we obtain a similar behavior.
 



\end{document}